\documentstyle[11pt]{article}
\newcommand{\re}{\mathop{\mathrm{Re}}}
\newcommand{\im}{\mathop{\mathrm{Im}}}
\begin{document}
\title{On the direct evidence of time-reversal non-invariance\\ in the
  $K^0-\bar K^0$ system}
\author{ Andr\'e Roug\'e\thanks{Andre.Rouge@in2p3.fr}\\
 LPNHE Ecole Polytechnique-IN2P3/CNRS\\
 F-91128 Palaiseau Cedex}
\date{August 1999}
\maketitle
\begin{abstract}
The measurements of the CP-violation parameters by the CPLEAR
experiment are reviewed. It is shown that attempts to prove T-violation
from the semileptonic asymmetries  are flawed by logical inconsistencies.
\end{abstract}
\mbox{~}\\[-10.5cm]
\begin{flushright}
X-LPNHE 99/05
\end{flushright}
\mbox{~}\\[7.5cm]
\section{Introduction}
The CPLEAR Collaboration has recently published a series of papers
\cite{cpl1,cpl2,cpl3,cpl4} reporting high precision measurements of
the parameters which describe the CP-violation in the $K^0-\bar K^0$ system. 
In the first paper \cite{cpl1}, the measurement of the semileptonic decay-rate asymmetry,
\begin{equation}
A_T(\tau)=\frac{R(\bar K^0_{t=0}\to e^+\pi^-\nu_{t=\tau})-
R( K^0_{t=0}\to e^-\pi^+\bar\nu_{t=\tau})
}{ R(\bar K^0_{t=0}\to e^+\pi^-\nu_{t=\tau})+
R( K^0_{t=0}\to e^-\pi^+\bar\nu_{t=\tau})}
\end{equation}
is presented. Assuming CPT-symmetry in the semileptonic decays and the
 $\Delta S=\Delta Q$ rule, the $A_T$ asymmetry is identified to the Kabir asymmetry
 \cite{Kabas},
\begin{equation}
A_K=\frac{{\cal P}(\bar K^0\to K^0)- {\cal P}( K^0\to\bar K^0)}{
 {\cal P}(\bar K^0\to K^0)+ {\cal P}( K^0\to\bar K^0)}
\end{equation}
and the measured value $A_T=(6.6\pm 1.64)\times 10^{-3}$ is
claimed to be a direct evidence for time-reversal non-invariance.

The validity of the identification of $A_T$ and $A_K$ has been
discussed from diverse points of view \cite{AKLP,ellis,PKK},  and it has been
asserted \cite{AKLP,ellis,PDV2} that its proof was established by a 
 subsequent measurement  \cite{cpl3}. We  discuss here this assertion.
\section{Semileptonic asymmetries}\label{sec:asym}
Let us first define the notations. The $K_S$ and $K_L$ states are
written
\begin{eqnarray}
K_S&=&K_1+\epsilon_S K_2\nonumber\\
K_L&=&K_2+\epsilon_L K_1
\end{eqnarray} 
in terms of the CP-eigenstates $K_1$ and $K_2$, and the
$\epsilon_{S,L}$ are decomposed into
\begin{equation}
\epsilon_S=\epsilon+\delta,~~~~~~~~~~\epsilon_L=\epsilon-\delta
\end{equation}
The violations of CP and  $\Delta S=\Delta Q$ in the semileptonic
decays are described by  the parameters $y$, $x_+$ and $x_-$
\cite{cpl1,dass,dalitz,buchanan,maiani}.
The constraints from  $\Delta S=\Delta Q$, CPT-, CP- and T-conservation 
are:
\begin{eqnarray}
\Delta S=\Delta Q&\hspace{2cm}&x_+=x_-=0\\
CPT&&\delta=y=x_-=0\\
CP&&\re(y)=\re(x_-)=\im(x_+)=0\\
T&&\re(\epsilon)=\im(y)=\im(x_+)=\im(x_-)=0
\end{eqnarray}
Only first order terms in the small parameters, $\epsilon$, $\delta$,
$x_\pm$ and $y$ are retained in the theoretical expressions.

Writing $\bar R_+=\bar R_+(\tau)$ instead of   $R(\bar K^0_{t=0}\to
e^+\pi^-\nu_{t=\tau})$, $\bar N_+$ the corresponding number of decays
and $\bar N$ the total number of $\bar K^0$, 
\begin{equation}
A_T=\frac{\bar R_+-R_-}{\bar R_++R_-}=\frac{\bar N_+/\bar N-N_-/N}{ \bar N_+/\bar N-N_-/N}
\end{equation}
In order to get a high precision determination of $\bar N/N$, the
CPLEAR Collaboration uses its measurements of the $\pi^+\pi^-$ decays. 
Since the decay-ratios must follow the law:
\begin{equation}\label{eq:pipidec}
R^{K0/\bar K^0}_{\pi^+\pi^-}=C ( 1\mp 2\re(\epsilon_L))
 [ e^{ -\Gamma_S\tau}+|\eta_{+-}|^2e^{-\Gamma_L \tau}
\pm 2|\eta_{+-}|e^{-\frac{1}{2}(\Gamma_S+\Gamma_L)\tau}
\cos (\delta m\tau-\phi_{+-}) ]
\end{equation}
it is required that the parameter $\re(\epsilon_L)$ determined from the
$\pi^+\pi^-$ decays is equal to $\delta_l/2$, where $\delta_l$ is the
measured \cite{PDG} charge asymmetry of the $K_L$. 
The constraint can be written $(1+\delta_l)\bar NN_{\pi\pi}=(1-\delta_l)N\bar
N_{\pi\pi}$ and the experimental asymmetry is:
\begin{equation}
A_T^{exp}(\tau)=\frac{N_{\pi\pi}\bar N_+(\tau)(1+\delta_l)-\bar  N_{\pi\pi}N_-(\tau)(1-\delta_l)}{
 N_{\pi\pi}\bar N_+(\tau)(1+\delta_l)+\bar  N_{\pi\pi}N_-(\tau)(1-\delta_l)}=
\frac{ N_{\pi\pi}\bar N_+(\tau)-\bar N_{\pi\pi}N_-(\tau)}{ N_{\pi\pi}\bar N_+(\tau)+\bar
  N_{\pi\pi}N_-(\tau)} +\delta_l
\end{equation}
neglecting second order in the CP-violation.

In a following paper \cite{cpl2}, the CPT-violating parameter 
$\re(\delta)$ is measured by the use of another asymmetry:
\begin{equation}
A_\delta(\tau)=A_1(\tau)+A_2(\tau)
\end{equation}
where the $A_1$ and $A_2$ asymmetries are defined by:
\begin{equation}
A_1(\tau)=\frac{ \bar R_+(\tau)-R_-(\tau)(1+4\re(\epsilon_L))}{
 \bar R_+(\tau)+R_-(\tau)(1+4\re(\epsilon_L))},\hspace{1cm}
A_2(\tau)=\frac{ \bar R_-(\tau)-R_+(\tau)(1+4\re(\epsilon_L))}{
 \bar R_-(\tau)+R_+(\tau)(1+4\re(\epsilon_L))}
\end{equation}
In order to get rid of the normalizations, $N$ and $\bar N$, the
parameter $\re(\epsilon_L)$ is extracted from the measured
$\pi^+\pi^-$ decay rates and the experimental expressions of $A_1$ and
$A_2$ are:
\begin{equation}
A_1(\tau)=\frac{ N_{\pi\pi}\bar N_+(\tau)-\bar N_{\pi\pi}N_-(\tau)}{N_{\pi\pi}\bar N_+(\tau)+\bar
  N_{\pi\pi}N_-(\tau)},\hspace{1cm}A_2(\tau)=\frac{ N_{\pi\pi}\bar N_-(\tau)-\bar N_{\pi\pi}N_+(\tau)}{ N_{\pi\pi}\bar N_-(\tau)+\bar
  N_{\pi\pi}N_+(\tau)}
\end{equation}
With the above notations, the $A_T^{exp}$ asymmetry is:
\begin{equation}
A_T^{exp}(\tau)=A_1(\tau)+\delta_l
\end{equation}

The theoretical expressions of the asymmetries, without any assumption
on CPT or $\Delta S=\Delta Q$, are \cite{cpl3,dass,dalitz,buchanan}:
\begin{eqnarray} 
A_1(\tau)&=&~~\,2\re(\epsilon-y-x_-)+2\re(\delta)+F_1(\tau;\im(x_+),\re(x_-))\\
A_2(\tau)&=&-2\re(\epsilon-y-x_-)+6\re(\delta)+F_2(\tau;\delta,
\im(x_+),\re(x_-))\\
\delta_l&=&~~\,2\re(\epsilon-y-x_-)-2\re(\delta )
\end{eqnarray}
with, for $\tau\gg\tau_S$, $F_1=F_2=0$.

The $\delta_l$ asymmetry and the time-independent parts of $A_1$ and
$A_2$ carry information on  two parameters only, $\re(\epsilon-y-x_-)$
and $\re(\delta)$. The $A_T^{exp}$ and $A_\delta$ asymmetries are
two of their possible estimators; the optimal ones can be obtained,
for instance, by a simultaneous fit of the three asymmetries.

The time-dependent parts bring information on
$\im(\delta)$, $\re(x_-)$  and $\im(x_+)$ and some additional information on
$\re(\delta)$. With the normalization procedures
used by  CPLEAR (see appendix) there is no independent information on $\epsilon$ and
$y$  in the decay-ratio asymmetries.
 
The proof of T-violation ($\re(\epsilon)\neq 0$)  from $A_T^{exp}$  requires a
measurement  of $\re(y+x_-)$. Such a measurement has been given by
CPLEAR \cite{cpl3} but, to be used, it must be independent of $A_T^{exp}$. 
It is easy to check that, on the contrary, 
 the measurement of $\re(y+x_-)$  is given by $A^{exp}_T$.   
\section{Global fit}
The determination of the CP-violation parameters in
\cite{cpl3} is made through a global fit 
 based on the Bell-Steinberger relation \cite{BS}:
\begin{eqnarray}\label{BSrel}
\lefteqn{
\re(\epsilon)-i\im(\delta)=
  {[}\Gamma_L+\Gamma_S+2i(m_L-m_S) {]}^{-1}\times}\hspace{13cm}\\\hbox{\large [}
\sum
\gamma_S^{\pi\pi}\eta_{\pi\pi}+\sum\gamma_L^{3\pi}\eta^*_{3\pi}
+2\gamma_L^{\pi
  l\nu}(\re(\epsilon-y)-i\im(x_++\delta))\,
\hbox{\large ]} \nonumber 
\end{eqnarray}
where $\gamma_{S,L}^f$ is the partial width of the $K_{S,L}$ in the
final state $f$.

The experimental data entered in the fit are the partial widths, the
$\eta$ amplitudes and also the semileptonic asymmetries, $A_1(\tau)$,
$A_2(\tau)$ and $\delta_l$. All the information used to measure
$A^{exp}_T$ and $A_\delta$ \cite{cpl1,cpl2} is also exploited, in an
optimal way, in the global fit.

As mentioned before the simultaneous fit of the semileptonic
asymmetries gives optimized  measurements of $\re(\delta)$ and
$\re(\epsilon-y-x_-)$, from the constant terms, of $\re(x_-)$ and
$\im(x_+)$ from the time-dependence. They can be used to compute the
quantity $\re(\epsilon-y)-i\im(x_+)$  appearing on the right-hand
side of the Bell-Steinberger relation which gives $\re(\epsilon)$ and $\im(\delta)$.  
However, the semileptonic contribution to eq. \ref{BSrel}
being tiny,  the parameters  
$\re(\epsilon)$ and $\im(\delta)$ are, in a very good approximation,
independent of the others and determined, as usual, by the
hadronic decays \cite{cpl3}. 

The only reason to include The Bell-Steinberger relation in a global
fit is to constrain the fit of the time-dependent part of $A_2$ by the
knowledge of $\im(\delta)$. The errors on the already measured
parameters can be slightly improved but there is no spontaneous
generation of information on new parameters.
Anyway, the value of the correlation
coefficient $C(\re(y),\re(x_-))=-0.997$ \cite{cpl3} shows that the
additional information from the time-dependence is very small.

So the parameter
$\re(\delta)$ is determined by the $A_\delta$ asymmetry, and
the parameter  $\re(\epsilon-y-x_-)$ by $A_T^{exp}$.  More
exactly, they are determined  by the
optimal combinations of $\delta_l$ and the time-independent parts of $A_1$
and $A_2$, constructed by the fit. The measurement of $\re(y+x_-)$ is
nothing else that the difference of  $\re(\epsilon-y-x_-)$ measured by
the optimized 
$A_T^{exp}$ and $\re(\epsilon)$ measured by the hadronic decays
through the Bell-Steinberger relation. Using it to correct the
measurement of  $A_T^{exp}$ for the possible CPT-violation in the
decay would just give back the value of $\re(\epsilon)$ provided by
the Bell-Steinberger relation.
\section{Conclusion} 
 In the very complete and precise set of measurements \cite{cpl1,cpl2,cpl3,cpl4} published
by the CPLEAR Collaboration, the  $A_T^{exp}$ asymmetry serves to measure
neither the T-violating parameter
$\re(\epsilon)$ \cite{AKLP} nor $\re(\delta)$ \cite{PKK} but 
the CPT-violating parameter $\re(y+x_-)$. There is no logical way to
use the measured $A_T^{exp}$ as a direct evidence of time-reversal
non-invariance  without assuming CPT-conservation in the semileptonic
decays. The proof of T-violation still rests on the Bell-Steinberger relation.
\section{Appendix}
In the following, each time-dependent asymmetry $A(\tau)$
 is decomposed
into a constant, asymptotic part $\tilde A$ and a time-dependent part
$F_a$:
\begin{equation}
A(\tau)=\tilde A+F_a(\tau),\hspace{1cm}F_a=0 \hbox{~~~for~~~} \tau\gg\tau_S
\end{equation}
From the four semileptonic decay-ratios, $R_+$, $R_-$, $\bar R_+$ and
$\bar R_-$, three independent asymmetries can be constructed. One is
the CP-conserving $A_{\delta m}$:
\begin{equation}
A_{\delta m}(\tau)=\frac{[R_+(\tau)+\bar R_-(\tau)]-[\bar R_+(\tau)+R_-(\tau)]}
{[R_+(\tau)+\bar R_-(\tau)]+[\bar R_+(\tau)+R_-(\tau)]}=
F_{\delta m}(\tau;\re(x_+),\im(x_-))
\end{equation} 
Assuming $\im(x_-)=0$, it has been used by CPLEAR \cite{cpl5} to
measure $|\delta m|$ and $\re(x_+)$.
The two others are CP-violating, for instance:
\begin{eqnarray}
A_T(\tau)&=&\frac{\bar R_+(\tau)-R_-(\tau)}{\bar
  R_+(\tau)+R_-(\tau)}=4\re(\epsilon) -2\re(y+x_-)+F_T(\tau)\\
A_{CPT}(\tau)&=&\frac{\bar R_-(\tau)-R_+(\tau)}{\bar
  R_-(\tau)+R_+(\tau)}=4\re(\delta)+2\re(y+x_-)+F_{CPT}(\tau)
\end{eqnarray} with $F_T=F_1$ and $F_{CPT}=F_2$.
Their asymptotic values give:
\begin{equation}
\tilde A_T+\tilde A_{CPT}=4\re(\epsilon_S),\hspace{1cm}\tilde
A_T-\tilde A_{CPT}=2\delta_l
\end{equation}
Since the measurement of the  $R_{\pi^+\pi^-}$ and $\bar
R_{\pi^+\pi^-}$ decay-ratios allows (eq. \ref{eq:pipidec}) a  
determination of $\re(\epsilon_L)=\re(\epsilon-\delta)$,
 the T-violating parameter $\re(\epsilon)$ can be
determined from single-channel decay-ratio measurements. 

The  determination of $\re(\epsilon)$ sketched above rests, however,
on the assumption that the total numbers of
$K^0$ and $\bar K^0$ are exactly known. If they are not, and the $\pi^+\pi^-$
decay-channel is used as normalization \cite{cpl1,cpl2,cpl3}, the $A_T$ and $A_{CPT}$
asymmetries are replaced by the previously defined $A_1=A_T-2\re(\epsilon_L)$ and  
$A_2=A_{CPT}-2\re(\epsilon_L)$ with
\begin{equation}
\tilde A_1+\tilde A_{2}=8\re(\delta),\hspace{1cm}\tilde
A_1-\tilde A_{2}=2\delta_l
\end{equation}
and the independent information from  $R_{\pi^+\pi^-}/\bar R_{\pi^+\pi^-}$ is
lost.

The determination of $\re(\delta)$, which is the greatest achievement of the
CPLEAR measurement of semileptonic decays, is still possible but
there is no more information in the remaining part of the asymptotic
asymmetries than in $\delta_l$.

The connection of $A_T^{exp}$ to a measurement of $\re(\delta)$
mentioned in \cite{PKK} can be described in a different way.
$A_T^{exp}$  is an estimator of $4\re(\epsilon-y-x_-)$ constructed
from $\re(\delta)$, given by the semileptonic asymmetries measured by
CPLEAR, and from $\delta_l$, given by a combination of CPLEAR
asymmetries and the PDG value. As stated before,
this estimator is not the best one; for instance,
 using simply the PDG value \cite{PDG} of
$\delta_l$ and the CPLEAR measurement \cite{cpl2} of $\re(\delta)$,
gives $4\re(\epsilon-y-x_-)=2\delta_l+\tilde A_\delta/2=(7.74\pm 1.36)\times 10^{-3}$.

\end{document}